\documentclass{LMCS}

\def\doi{9(1:01)2013}
\lmcsheading%
{\doi}
{1--27}
{}
{}
{Jun.~16, 2011}
{Feb.~14, 2013}
{}

\usepackage{includes}
\usepackage[hyperfootnotes=false]{hyperref}
\usepackage{enumerate}

\begin{document}

\title[Type classes for efficient exact real arithmetic in Coq]{Type classes for efficient exact \\ real arithmetic in Coq}
\author[R.~Krebbers]{Robbert Krebbers}
\author[B.~Spitters]{Bas Spitters}
\address{Institute for Computing and Information Sciences\\
	Radboud University Nijmegen}
\email{mail@robbertkrebbers.nl, spitters@cs.ru.nl}
\thanks{The research leading to these results has received funding from the European Union's 7th Framework Programme under grant agreement nr.~243847 (ForMath).}

\begin{abstract}
Floating point operations are fast, but require continuous effort by
the user to ensure correctness. This burden can be shifted to the
machine by providing a library of \emph{exact} analysis in which the
computer handles the error estimates. Previously, we provided a fast
implementation of the exact real numbers in the \Coq{} proof
assistant. This implementation incorporates various optimizations to
speed up the basic operations of O'Connor's implementation by a 100
times. We implemented these optimizations in a modular way using type
classes to define an abstract specification of the underlying dense
set from which the real numbers are built. This abstraction does not
hurt the efficiency.

This article is a substantially expanded version of
(Krebbers/Spitters, Calculemus 2011) in which the implementation is
extended in the various ways. First, we implement and verify the sine
and cosine function. Secondly, we create an additional implementation
of the dense set based on \Coq's fast rational numbers. Thirdly, we
extend the hierarchy to capture order on undecidable structures, while
it was limited to decidable structures before. This hierarchy, based
on type classes, allows us to share theory on the naturals, integers,
rationals, dyadics, and reals in a convenient way. Finally, we obtain
another dramatic speed-up by avoiding evaluation of termination proofs
at runtime.
\end{abstract}

\ACMCCS{[{\bf Theory of Computation}]: Logic---Type theory \&
             Constructive mathematics;
    [{\bf Mathematics of Computing}]: Mathematical analysis---Numerical analysis---Arbitrary-precision arithmetic}
\maketitle

\section{Introduction}
There is a big gap between numerical algorithms in research papers, which typically use concepts like Hilbert spaces and fixed point theorems from functional analysis, and their actual implementation, which uses floating point numbers\footnote{By floating points we mean numbers of the shape $n * b ^ e$, where $n$ and $e$ are bounded integers and $b$ is the base for scaling (typically 2, 10 or 16). The most widely used form of floating point arithmetic is the IEEE 754 standard, which is present in many hardware and software implementations.} and matrix operations. 
This gap makes it difficult to trust the code. Similarly, this gap is undesirable in proofs of theorems (e.g.\ Kepler conjecture~\cite{hales02verification}, existence of the Lorentz attractor~\cite{tucker2002rigorous}) that rely on the execution of this code for numerical approximation. Finally, from a purely software engineering point of view, the situation is undesirable, because the gap between the (abstract mathematical) numerical algorithms and the (concrete floating point) implemented program makes the code difficult to maintain.

The challenge to close this gap has already been posed by Bishop in his fundamental work on constructive analysis~\cite{Bishop67}. Bishop proposed to use constructive analysis to bridge this gap. Moreover, we can narrow this gap by using
\begin{iteMize}{$\bullet$}
\item exact real numbers or intervals instead of floating point numbers;
\item functional programming instead of imperative programming;
\item dependent type theory which allows us to compute with complete metric spaces;
\item a proof assistant which allows us to verify the correctness proofs;
\item constructive mathematics to tightly connect mathematics with computations and to avoid computationally impossible case distinctions.
\end{iteMize}
Separately, all these tools have proved itself. By going to the limits of this proven technology we should be able to come within a small constant factor of floating point computations. In this way one would obtain a tool suitable for research and education in numerical analysis that allows one to compute abstractly at the level of functional analysis, e.g.\ to compute fixed points of operators on Hilbert spaces. Like the development of \Name{Fortran} and \Name{MATLAB} this will require a huge amount of work. In the present paper we focus on the performance of real number computation in the \Coq{} proof assistant.

Real numbers, being infinite objects, cannot be represented exactly in a computer. Hence, in constructive analysis~\cite{Bishop67} one uses functions which when fed a desired precision approximate a real numbers by a rational, or a dyadic number, to within that precision\footnote{One could argue that we capture only the definable, or computable, real numbers in this way. These issues are important and well-studied, see for instance~\cite{troelstra1988constructivism}, but we will not go into them here.}.
The real numbers are the completion of the rationals. This completion construction can be organized in a monad, a familiar construct from functional programming. The completion monad provides an efficient combination of proving and computing~\cite{OConnor:mscs}. In this way, O'Connor~\cite{Oconnor:real} implements exact real numbers and the transcendental functions on them in \Coq. A number of possible improvements in this implementation were already suggested in~\cite{Riemann,oconnor-thesis}.
\begin{enumerate}[(1)]
\item Use \Coq's new machine integers instead of the integers built from ordinary inductive data types;
\item use dyadic rationals (that are numbers of the shape $n * 2 ^ e$ for $n,e \in \Z$, also known as infinitary floats) instead of ordinary rationals;
\item use approximate division to improve the implementation of power series.
\end{enumerate}
Here we carry out all three optimizations. Unfortunately, changing O'Connor's implementation to use the new machine integers was far from trivial, as he used a particular concrete representation of the rationals. To avoid this in the future, we moreover provide an abstract specification of the dense set as \emph{approximate rationals}. Finally, we obtain another dramatic speed-up by avoiding evaluation of termination proofs at runtime.

\subsection*{Outline}
Section~\ref{section:coq} describes some aspects of the \Coq{} proof assistant relevant for our development. Section~\ref{section:metricspaces} describes metric spaces, monads, and O'Connor's implementation of the real numbers~\cite{OConnor:mscs}. Section~\ref{section:interfaces} extends Spitters and van der Weegen's approach to abstract interfaces using type classes~\cite{math-classes}. Section~\ref{section:reals} describes the theory of approximate rationals, our implementation of the real numbers, and deals with computing power series and the square root. We finish with some benchmarks in Section~\ref{section:bench} and conclusions in Section~\ref{section:conclusions}. The sources of our developments can be found at~\url{https://github.com/c-corn/corn}.

\section{The \Coq-system}
\label{section:coq}
The \Coq{} proof assistant is based on the calculus of inductive constructions~\cite{CoquandHuet,CoquandPaulin}, a dependent type theory with (co)inductive types; see~\cite{Coq,BC04}. In true Curry-Howard fashion, it is both a pure functional programming language with an expressive type system, and a language for mathematical statements and proofs. We highlight some aspects of \Coq{} relevant for our development.

\subsection{Notations}
\Coq{} has an extensible mechanism for defining complex notations.
We use this mechanism heavily, together with unicode symbols, to obtain notations that are closer to common mathematical practice.
However, due to conflicts with standard \Coq{} syntax, there are some small deviations. For example, we write $\forall x, P x$ instead of $\forall x. P x$.
In this paper we tried to stay as close as possible to the notations in our \Coq{} development.

\subsection{Types and propositions.}
\label{section:type_prop}
Propositions in \Coq{} are types~\cite{ITT,MartinLof:1982}, which themselves have types called \emph{sorts}. \Coq{} features a distinguished sort called \Prop\ that one may choose to use as the sort for types representing propositions. The distinguishing feature of the \Prop\ sort is that terms of non-\Prop\ type may not depend on the values of inhabitants of \Prop\ types (that is, proof terms). This regime of discrimination establishes a weak form of proof irrelevance, in that changing a proof can never affect the result of value computations. On a practical level, this lets \Coq{} safely erase all \Prop\ components when extracting certified programs to \OCaml{} or \Haskell. We should note however, that in practice, \Coq's extraction mechanism~\cite{letouzey2008extraction} is still very hard to use for programs with the complexity, in terms of depth of definitions, that we are interested in \cite{cruzfilipe2003,cruzfilipe2006}.

\subsection{Constructive indefinite description}
\label{section:indefinitive}
In spite of the restriction on \Prop{} discussed in the previous paragraph, \Coq{} allows recursive functions to use a value of \Prop\ type to ensure termination~\cite[14.2.3, 15.4]{BC04}.
In particular, this is used to prove \emph{constructive indefinite description}, which states that given a decidable predicate over the natural numbers, a \Prop{} based existential can be converted into a \Type{} based one. Its formal statement can be found in the standard library:
\begin{lstlisting}
Lemma constructive_indefinite_description_nat (P : nat → Prop) :
  (∀ x : nat, {P x} + {¬P x}) → (∃ n : nat, P n) → {n : nat | P n}
\end{lstlisting}
Here the notation $\{ x : A \separator P\ x \}$ for $P : A → \Prop$ denotes a Σ-type. This lemma can be seen as a form of Markov's principle in \Coq. The algorithm does a bounded search for a new witness satisfying the predicate. The witness from the \Prop{} based existential is only used to prove termination of the search. No information flows from the \Prop{} universe to the \Type{} universe because the witness found for the \Type{} based existential is independent of the witness from the \Prop{} based one.

\subsection{Equality, setoids, and rewriting}
Because the \Coq{} type theory lacks quotient types (as it endangers the decidability of type checking), one usually bases abstract structures on a \emph{setoid} (`Bishop set'): a type equipped with an equivalence relation~\cite{Bishop67,Hofmann}. This leads to a naive set theory as described by Palmgren~\cite{palmgren2009constructivist}. When the user attempts to substitute a given (sub)term using an equality, the system keeps track of, resolves, and combines proofs of equivalence~\cite{Setoid-rewrite}.

The `native' notion of equality in \Coq{}, \emph{Leibniz equality}, is that of terms being convertible, naturally reified as a proposition by the inductive type family \lstinline|eq| with single constructor \lstinline|eq_refl : ∀ (T : Type) (x : T), x ≡ x|, where \lstinline|a ≡ b| is notation for \mbox{\lstinline|eq T a b|}. Since convertibility is a congruence, a proof of \lstinline|a ≡ b| lets us substitute \lstinline|b| for \lstinline|a| anywhere inside a term without further conditions. Our interest is in more complicated equalities, so we diverge from \Coq{} tradition and reserve \lstinline|=| for setoid equality.
Rewriting with \lstinline|=| \emph{does} give rise to side conditions. For instance, consider formal fractions of integers as a representation of rationals. Rewriting a subterm using such an equality is permitted only if the subterm is an argument of a function that has been proven to \emph{respect} the equality. Such a function is called \lstinline|Proper|, and that property must be proved for each function in whose arguments we wish to enable rewriting.

\subsection{Type classes}
Type classes are a great success in the \Haskell{} functional programming language, as a means of organizing interfaces of abstract structures. \Coq's type classes provide a superset of their functionality, but are implemented in a different way.

In \Haskell{} and \Isabelle, type classes and their instances are second class. They are handled as specialized syntactic constructs whose semantics are given specifically by the type class apparatus. By contrast, the expressivity of dependent types and inductive families as supported in \Coq, combined with the use of pre-existing technology in the system (namely proof search and implicit arguments) enable a \emph{first class} type class implementation~\cite{DBLP:conf/tphol/SozeauO08}: classes are ordinary record types (`dictionaries'), instances are ordinary constants of these record types (registered as \emph{hints} with the proof search machinery), class constraints are ordinary implicit arguments, and instance resolution is achieved by augmenting the unification algorithm to invoke ordinary proof search for implicit arguments of class type.
Thus, type classes in \Coq{} are realized by relatively minor syntactic aids that bring together existing facilities of the theory and the system into a coherent idiom, rather than by introduction of a new category of qualitatively different definitions with their own dedicated semantics.

We use the algebraic hierarchy based on type classes and its abstract specification of $\N,\Z$ and $\Q$ described in~\cite{math-classes}. Unfortunately, we should note that we have clearly met the efficiency problems connected to the current implementation of type classes in \Coq. Luckily, these efficiency problems are limited to instance resolution which is only performed at compile time. Type classes effect the computation time of type checked terms due to the absence of code inlining. In an illustrative example the use of type classes caused only a 3\% performance penalty; see Section~\ref{section:bench}. 

\subsection{Virtual machine and machine integers}
\label{section:machine}
\Coq{} includes a virtual machine~\cite{Compiler}, \mbox{\lstinline|vm_compute|,} based on \OCaml{}'s virtual machine to allow efficient evaluation. 
Both the abstract machine and its compilation scheme have been proved correct, in \Coq, with respect to the weak reduction semantics. However, we still need to extend our trusted core to a bigger kernel, as the \emph{implementation} has not been formally verified.

Machine integers were also added to the \Coq{} system~\cite{machineintegers}. The usual evaluation inside \Coq{} (\lstinline|compute|) uses a special inductive type for cyclic integers, but the virtual machine (\lstinline|vm_compute|) uses actual machine integers.
The type \lstinline|bigZ| of arbitrary precision integers is built from binary trees of these cyclic integers.
Primality tests in~\cite{thesisSpiwack} show a big speed-up compared to the inductively defined integers.
Our work confirms this big speed-up gained by using machine integers.
We pay for this speed-up, however, by having to trust the virtual machine and its translation to actual machine integers. 

\section{Metric spaces and the completion monad}
\label{section:metricspaces}
Having completed our brief description of the \Coq-system, we now turn to O'Connor's formalization of exact real numbers~\cite{OConnor:mscs}. Traditionally, a metric space is defined as a set $X$ with a metric function $d:X \times X → \nonneg\R$ satisfying certain axioms. We use a more relaxed definition of a metric space that does not require the metric be a function; see also~\cite{Richman:2008}. The metric is represented via a (respectful) ball relation $\ballsym: \pos\Q → X → X → \Prop$ satisfying:\footnote{We use the positive rational numbers $\pos\Q$ instead of the non-negative relation numbers $\nonneg\Q$ as it can be expressed that two points $x$ and $y$ are within $0$ of each other by $∀ ε,\ \ball {ε} x y$.}
\begin{lstlisting}
  msp_refl : $∀ x\,ε,\ \ball {ε} x x$
  msp_sym : $∀ x\,y\,ε,\ \ball {ε} x y → \ball {ε} y x$
  msp_triangle : $∀ x\,y\,z\,ε_1\,ε_2,\ \ball {ε_1} x y → \ball {ε_2} y z → \ball {ε_1 + ε_2} x z$
  msp_closed : $∀ x\,y\,ε,\ (∀ δ,\ \ball {ε + δ} x y) → \ball {ε} x y$
  msp_eq : $∀ x\,y,\ (∀ ε,\ \ball {ε} x y) → x = y$
\end{lstlisting}
The ball relation $\ball{ε} x y$ expresses that the points $x$ and $y$ are within $ε$ of each other. We call this a ball relationship because the partially applied relation $\ballsym^X_{ε}\,x : X → \Prop$ is a predicate that represents the closed ball of radius $ε$ around the point $x$. For example, the ball relation on $\Q$ is $\ball[\Q] {ε} x y := |x - y| ≤ ε$.

A metric space $X$ is a \emph{prelength} space if:
\[
  ∀ a\,b\,ε\,δ_1\,δ_2,\ ε < δ1 + δ2 → \ball {ε} a b → ∃ c,\ \ball {δ_1} a c\ ∧\ \ball {δ_2} c b.
\]
In particular, a prelength space has approximate midpoints: for any $δ>0$ we can take $δ_1=δ_2=ε/2+δ$. Every complete prelength metric space is a length metric space. The metric space $\Q$ is a prelength space; see~\cite{OConnor:mscs} for details.

We will introduce the completion of a metric space as a monad. In order to do this we will first introduce monads.

\subsection{Monads} Moggi~{\cite{moggi:1989}} recognized that many non-standard forms of computation may be modeled by monads\footnote{In category theory one would speak about the Kleisli category of a (strong) monad.}. Wadler~{\cite{Wadler92a}} popularized their use in functional programming. Monads are now an established tool to structure computation with side-effects. For instance, programs with input $X$ and output $Y$ which have access to a mutable state $S$ can be modeled as functions of type $X \times S → Y \times S$, or equivalently $X → (Y \times S)^S$. The type constructor $\monad Y := (Y \times S)^S$ is an example of a monad. Similarly, partial functions may be modeled by maps $X → Y_{\bot}$, where $Y_{\bot} := Y + ()$ is a monad.

The formal definition of a (strong) monad is a triple $(\monad, \return, \bind)$ consisting of a type
constructor $\monad$ and two functions:
\begin{flalign*}
   \return &: X → \monad X \\
   \bind   &: (X → \monad Y) → \monad X → \monad Y
\end{flalign*}
We will denote $\return x$ as $\returnShort x$, and $\bind f$ as $\bindShort f$. These two operations must satisfy:
\begin{flalign*}
  \bind\ \return\ a &= a \\
  \bindShort f\,  \returnShort a  &= f\, a \\
  \bindShort f\, (\bindShort g\, a) &= \bind\ (\bindShort f ∘ g)\, a
\end{flalign*}
\subsection{Completion monad}
\label{section:completion-monad}
The completion of a metric space $X$ is defined by:
\[ 
	\complete X := \{ f : \pos\Q → X \separator ∀ ε_1\,ε_2,\ \ball {ε_1 + ε_2} {(f\,ε_1)} {(f\,ε_2)} \}.
\]
Given metric spaces $X$ and $Y$, a function $f : X → Y$ is \emph{uniformly continuous} with \emph{modulus} $μ_f : \pos\Q → \pos\Q$ if:
\[
	∀ ε\,x_1\,x_2,\ \ball {μ_f ε} {x_1} {x_2} → \ball {ε} {(f\,x_1)} {(f\,x_2)}.
\]
Completion is a monad on the category of metric spaces with uniformly continuous functions. The function $\return : X → \complete X$ defined by $λ x\, ε,\,x$ is the embedding of a metric space in its completion. Moreover, a uniformly continuous function $f : X → \complete Y$ with modulus $μ_f$ can be lifted to operate on complete metric spaces as $\bind f : \complete X → \complete Y$ defined by $λ x\,ε,\, f\, (x\, (µ_f \frac{ε}{2}))\, \frac{ε}{2}$. A restriction to prelength spaces is essential for this efficient definition of \bind; see~\cite{OConnor:mscs} for details.

One advantage of this approach is that it helps us to work with simple representations. Let $\R := \complete \Q$. Then to specify a function from $\R → \R$, we define a uniformly continuous function $f : \Q → \R$, and obtain $\bindShort f : \R → \R$ as the required function. Hence, the completion monad allows us to do in a structured way what was already folklore in constructive mathematics: to work with simple, often decidable, approximations to continuous objects.

\section{Abstract interfaces using type classes}
\label{section:interfaces}
An important part of this work is to further develop the algebraic hierarchy based on type classes by Spitters and van der Weegen~\cite{math-classes}. Especially, we extend their hierarchy with constructive fields, order theory and interfaces for mathematical operations, such as shift and power, common in programming languages. This layer of abstraction makes both proof engineering and programming more flexible: it avoids duplication of code, it introduces a canonical way to refer to operations and properties, both by names and notations, and it allows us to easily swap different implementations of number representations and their operations. First we will briefly recap the design decisions made in~\cite{math-classes}. For a nice tutorial following this design see~\cite{casteran:hal-00702455}. More information on type classes and setoids in \Coq{} can also be found in the reference manual~\cite{Coq}.

Algebraic structures are expressed in terms of a number of carrier sets, a number of relations and operations, and a number of laws that these operations and relations must satisfy. One way of describing such a structure is by a \emph{bundled representation} (as used in~\cite{FTA} for example): one uses a dependently typed record that contains the carrier, operations and laws. A setoid can be represented as follows.
\begin{lstlisting}
Record Setoid : Type := {
  st_car :> Type;
  st_equiv : st_car → st_car → Prop;
  st_setoid : Equivalence st_eq }.
Infix "=" := st_equiv : type_scope.
\end{lstlisting}
The notation \lstinline|:>| registers the projection \lstinline|st_car : Setoid → Type| as a coercion. Using the above interface for \lstinline|Setoid|s, one can now define a \lstinline|SemiGroup| whose carrier is a setoid.
\label{lstlisting:semigroup_bundled}
\begin{lstlisting}
Record SemiGroup : Type := { 
  sg_car :> Setoid;
  sg_op : sg_car → sg_car → sg_car;
  sg_proper : Proper (st_equiv ==> st_equiv ==> st_equiv) sg_op;
  sg_ass : ∀ x y z, sg_op x (sg_op y z) = sg_op (sg_op x y) z) }
\end{lstlisting}
The field \lstinline|sg_proper| states that the operation \lstinline|sg_op| respects the setoid equality. Its definition expands to \lstinline|∀ x1 x2, x1 = x2 → ∀ y1 y2, y1 = y2 → sg_op x1 y1 = sg_op x2 y2|.

However, this approach has some serious limitations, the most important one being a lack of support for \emph{sharing} components. For example, suppose we want to group together two \lstinline|CommutativeMonoid|s in order to create a \lstinline|SemiRing|. Now awkward hacks are necessary to establish equality between the carriers\footnote{An elegant solution is proposed by Pollack~\cite{Pollack:2002}. However, its implementation requires simultaneous inductive recursive definitions which are currently not supported in \Coq.}. A second problem is that if we stack up these records to represent higher structures the projection paths become increasingly long. In the above example, the projection path to obtain the carrier of a semigroup \lstinline|G| is \lstinline|st_car (sg_car G)|, but for fields, this path will be much longer.

Historically these problems have been an acceptable trade-off because \emph{unbundled representations}, in which the carrier and operations are parameterized, introduce even more problems.
An unbundled representation of a semigroup is as follows.
\begin{lstlisting}
Record SemiGroup A (Ae : A → A → Prop) (Aop : A → A → A) : Prop := {
  sg_setoid : Equivalence Ae;
  sg_op_proper : Proper (Ae ==> Ae ==> Ae) Aop;
  sg_ass : ∀ x y z, Ae (Aop x (Aop y z)) (Aop (Aop x y) z) }
\end{lstlisting}
There is nothing to bind notation to, no structure inference, and declaring and passing requires too much manual bookkeeping. Spitters and van der Weegen have proposed a use of \Coq's new type class machinery that resolves many of the problems of unbundled representations. Our current experiment confirms that this is a viable approach. 

An alternative solution is provided by packed classes~\cite{packed} which use an alternative, and older, implementation of a variant of type classes: canonical structures; see also Section~\ref{section:conclusions}. Yet another approach would be to use modules. However, as these are not first class, we would be unable to define, e.g.\ homomorphisms between algebraic structures.

The first step of the approach of Spitters and van der Weegen is to define an \emph{operational type class} for each operation and relation.
\begin{lstlisting}
Class Equiv A := equiv: relation A.
Infix "=" := equiv: type_scope.
Notation "(=)" := equiv (only parsing).
Class SgOp A := sg_op: A → A → A.
Infix "&" := sg_op (at level 50, left associativity).
Notation "(&)" := sg_op (only parsing).
\end{lstlisting}
\Haskell-style notations \lstinline|(=)| and \lstinline|(&)| are defined so operations and relations can easily be used in partially applied position.

Now an algebraic structure is just a type class living in \Prop{} that is parametrized by its carrier, relations and operations. This class contains all laws that the operations should satisfy. The class for semigroups is as follows\footnote{We sometimes use the \lstinline|@| prefix to bypass implicit arguments in order to avoid ambiguity.}.
\begin{lstlisting}
Class Setoid A {Ae : Equiv A} : Prop := setoid_eq :> Equivalence (@equiv A Ae).
Class SemiGroup A {Ae : Equiv A} {Aop: SgOp A} : Prop := {
  sg_setoid :> @Setoid A Ae;
  sg_op_proper :> Proper ((=) ==> (=) ==> (=)) (&);
  sg_ass :> Associative (&) }.
\end{lstlisting}   
Since the operations are unbundled we can easily support sharing. First we make classes for the semiring operations and show that these are in fact special instances of the group operations. For example\footnote{We use \lstinline|(.*.)| instead of \lstinline|(*)| due to conflicting notations with \Coq's comments.}:
\begin{lstlisting}
Class Mult A := mult: A → A → A.
Infix "*" := mult.
Notation "(.*.)" := mult (only parsing).
Instance mult_is_sg_op `{f : Mult A} : SgOp A := f.
\end{lstlisting}
The \lstinline|SemiRing| class is then as follows.
\begin{lstlisting}
Class SemiRing A {Ae : Equiv A} {Aplus : Plus A} 
    {Amult : Mult A} {Azero : Zero A} {Aone : One A} : Prop := { 
  semiplus_monoid :> @CommutativeMonoid A Ae plus_is_sg_op zero_is_mon_unit;
  semimult_monoid :> @CommutativeMonoid A Ae mult_is_sg_op one_is_mon_unit;
  semiring_distr :> LeftDistribute (.*.) (+);
  semiring_left_absorb :> LeftAbsorb (.*.) 0 }.
\end{lstlisting}
The syntax \lstinline|:>| in the definition of \lstinline|SemiRing| declares certain fields as substructures\footnote{This syntax should not be confused with the similar syntax for coercions in records (e.g. in the bundled representation of a \lstinline|SemiGroup| on page~\pageref{lstlisting:semigroup_bundled}).},
so that in any context where $(A, =, +, *, 0, 1)$ is known to be a \lstinline|SemiRing|, $(A, =, +, 0)$ and $(A, =, ∗, 1)$ are automatically known to be \lstinline|CommutativeMonoid|s (and so on, transitively, because instance resolution is recursive). In our hierarchy, these substructures by themselves establish the inheritance diagram as in Figure~\ref{figure:hierarchy}.

Without type classes it would be cumbersome to manually carry around the arguments of the class. However, because these arguments are type classes themselves, the type class machinery will perform that job for us. Therefore, all arguments, except the carrier \lstinline|A| are declared as implicit using the syntax \lstinline|{x : X}|, so the user does not have to specify them.

Proving that an actual structure is an instance of the \lstinline|SemiRing| interface is straightforward. First we define instances of the operational type classes.
\begin{lstlisting}
Instance nat_equiv: Equiv nat := eq.
Instance nat_plus: Plus nat := Peano.plus.
Instance nat_mult: Mult nat := Peano.mult.
Instance nat_0: Zero nat := 0%nat.
Instance nat_1: One nat := 1%nat.
\end{lstlisting}
Here we see that instances are just ordinary constants of the class types. However, we use the \lstinline|Instance| keyword instead of \lstinline|Definition| to let the type class machinery register the instance. Now, to prove that the Peano naturals are in fact a semiring, we just write:
\begin{lstlisting}
Instance: SemiRing nat.
Proof.   ...   Qed.
\end{lstlisting}
The implicit arguments of \lstinline|SemiRing nat| are automatically inferred by instance search. In order to type check \lstinline|SemiRing nat|, it has to solve \lstinline|@SemiRing nat ?1 ?2 ?3 ?4 ?5| with obligations \lstinline|?1 : Equiv nat|, \ldots, \lstinline|?5 : One nat|. Since we have declared instances \lstinline|nat_equiv : Equiv nat|, \ldots, \lstinline|nat_1 : One nat|, type class search will trivially solve these obligations. Thus \lstinline|SemiRing nat| is actually \lstinline|@SemiRing nat nat_equiv nat_plus nat_mult nat_0 nat_1| with all type class constraints resolved.

The \lstinline|SemiRing| type class can be used as follows.
\begin{lstlisting}
Lemma example `{SemiRing A} x : 1 * x = x + 0.
\end{lstlisting}
The backtick instructs \Coq{} to automatically insert implicit declarations, namely \lstinline|Ae Aplus Amult Azero Aone|. It also lets us omit a name for the \lstinline|SemiRing A| argument itself. All of these arguments will be given automatically generated names that we will never refer to. Furthermore, instance resolution will automatically find instances of the operational type classes for the written notations. Thus the above is really:
\begin{lstlisting}
Lemma example {A Ae Aplus Amult Azero Aone} {P : @SemiRing A Ae Aplus Amult Azero Aone} (x : A) : 
  @equiv A Ae
    (@mult A Amult (@one A Aone) x) 
    (@plus A Aplus x (@zero A Azero)).
\end{lstlisting}

This approach to interfaces proved useful to formalize a standard algebraic hierarchy. Combined with category theory and universal algebra, $\N$ and $\Z$ are represented as interfaces specifying an initial semiring and initial ring~\cite{math-classes}.
\begin{lstlisting}
Class NaturalsToSemiRing (A : Type) :=
  naturals_to_semiring : ∀ B `{Mult B} `{Plus B} `{One B} `{Zero B}, A → B.  
Class Naturals A {Ae Aplus Amult Azero Aone} `{U : NaturalsToSemiRing A} :=  { 
  naturals_ring :> @SemiRing A Ae Aplus Amult Azero Aone;
  naturals_to_semiring_mor :> ∀ `{SemiRing B}, SemiRing_Morphism (naturals_to_semiring A B);
  naturals_initial :> Initial (semirings.object A) }.
\end{lstlisting}
These abstract interfaces for the naturals and integers make it easy to change the concrete representation in the future.
As fields are not algebraic, no such algebraic specification exists for the rational numbers. Hence, we choose to specify $\Q$ as the field of fractions of \Z. More precisely, $\Q$ is specified as a field containing $\Z$ that moreover can be embedded into the field of fractions of $\Z$.
\begin{lstlisting}
Inductive Frac A {Ae : Equiv A} {Azero : Zero A} : Type := 
  frac { num : A; den : A; den_ne_0 : den ≠ 0 }.
Class RationalsToFrac (A : Type) := rationals_to_frac : ∀ B `{Integers B}, A → Frac B.
Class Rationals A {Ae Aplus Amult Azero Aone Aneg Arecip} `{U : !RationalsToFrac A} : Prop :=  { 
  rationals_field :> @DecField A Ae Aplus Amult Azero Aone Aneg Arecip; 
  rationals_frac :> ∀ `{Integers Z}, Injective (rationals_to_frac A Z); 
  rationals_frac_mor :> ∀ `{Integers Z}, SemiRing_Morphism (rationals_to_frac A Z); 
  rationals_embed_ints :> ∀ `{Integers Z}, Injective (integers_to_ring Z A) }.
\end{lstlisting}

In current versions of \Coq{}, inference of substructures is based on \emph{backward} reasoning. In our semiring example that means, each time a \lstinline|CommutativeMonoid A| instance is needed, instance search may try to find a \lstinline|SemiRing A| instance. This style of instance search presents some problems, as the following example illustrates.
\begin{lstlisting}
Class Setoid_Morphism {A B} {Ae : Equiv A} {Be : Equiv B} (f : A → B) := { 
  setoidmor_a :> Setoid A;
  setoidmor_b :> Setoid B;
  sm_proper :> Proper ((=) ==> (=)) f }.
\end{lstlisting}
Each time we have to establish \lstinline|Setoid R| for some \lstinline|R|, instance search might try to infer a \lstinline|Setoid_Morphism| from an arbitrary \lstinline|S| to \lstinline|R|, or vice versa. Since this search quickly results in a serious blow-up, we omit the substructure declaration \lstinline|:>|. Support for \emph{forward} reasoning may solve this problem. If we would be in a context in which we know something to be a \lstinline|Setoid_Morphism|, then forward reasoning automatically infers that the source and target are \lstinline|Setoid|s. Recently, an initial implementation of forward reasoning has been added to \Coq{}, but it suffers from some other performance problems.

\subsection{Constructive fields and apartness}
\label{section:apartness}
In constructive mathematics, the common notion of inequality as the negation of equality is often too weak because a proof of a negation lacks computational content. For example, in order to define the reciprocal of $x \in \R$, one needs a witness $ε \in \pos\Q$ that $|x|\geq ε$. Such a witness cannot be extracted from a proof of $x ≠ 0$. To solve this problem, one uses a setoid equipped with an apartness (irreflexive, asymmetric and co-transitive) relation describing inequality~\cite{troelstra1988constructivism}.

The algebraic hierarchy in the CoRN library~\cite{C-corn} has been built on top of such setoids. 
Unfortunately, this hierarchy is quite `heavy' in practice. First, for structures with decidable equality, the negation of equality is the only \emph{tight} apartness. Hence, when working with decidable structures, an apartness relation is unnecessary. Secondly, CoRN uses an \emph{informative} (that is, \Type{} based) apartness relation to facilitate extraction of witnesses. However, \Coq's present implementation of setoid rewriting 
does not support rewriting over relations in \Type{}. So, it does not allow us to replace equations in expressions involving CoRN's informative apartness and thus many proofs involve a lot of manual labor.

To remedy these issues we propose an alternative solution. We use a \emph{non-informative} (that is, \Prop-based) apartness relation to enable setoid rewriting and include it just in the parts of the algebraic hierarchy where we actually need it. The latter keeps our interfaces clean and easy to use and should combine the best of two worlds. Type classes are of great help to reduce bookkeeping and clutter in proofs.

Although using a non-informative apartness relation enables setoid rewriting, it disables extraction of witnesses. Fortunately, in case of the reals, a witness can be obtained inefficiently by bounded linear search (see Section~\ref{section:indefinitive} and \ref{section:reals_order}). We think our approach is a reasonable trade-off since the amount of reasoning exceeds the potential use of apartness in computation. In case we need a witness for efficient computation, we just have to specify it explicitly. This approach of specifying witnesses explicitly was already preferred by O'Connor~\cite{Oconnor:real}, even when an informative apartness was available.

Our interface for a setoid with apartness (henceforth \lstinline|StrongSetoid|) is as follows.
\begin{lstlisting}
Class Apart A := apart: relation A.
Infix "⪥" := apart (at level 70, no associativity) : type_scope.

Class StrongSetoid A {Ae: Equiv A} {Aap : Apart A} : Prop :=  { 
  strong_setoid_irreflexive :> Irreflexive (⪥);
  strong_setoid_symmetric :> Symmetric (⪥);
  strong_setoid_cotrans :> CoTransitive (⪥);
  tight_apart : ∀ x y, ¬x ⪥ y ↔ x = y }.
\end{lstlisting}
This interface is equipped with a \emph{tight} equality. We prove that each \lstinline|StrongSetoid| is a \mbox{\lstinline|Setoid|.} For decidable structures, we define the following class to describe that the apartness relation is the negation of equality.
\begin{lstlisting}
Class TrivialApart A `{Equiv A} {ap : Apart A} := trivial_apart : ∀ x y, x ⪥ y ↔ x ≠ y.
\end{lstlisting}
Given a setoid with decidable equality we can easily extend it to a \lstinline|StrongSetoid|.
\begin{lstlisting}
Instance default_apart `{Equiv A} : Apart A | 20 := (≠).
Instance default_apart_trivial `{Equiv A} : TrivialApart A (ap:=default_apart).
Lemma dec_strong_setoid `{Setoid A} `{Apart A} 
  `{!TrivialApart A} `{∀ x y, Decision (x = y)} : StrongSetoid A.
\end{lstlisting}
Unfortunately, the type class mechanism is unable to detect simple loops. Hence we define \lstinline|dec_strong_setoid| as an ordinary \lstinline|Lemma| instead of an \lstinline|Instance|. This trick prevents \Coq{} from using it in instance search and therefore avoids endless derivations of the form \lstinline|StrongSetoid A|, \lstinline|Setoid A|, \lstinline|StrongSetoid A|, \ldots

For ordinary setoids we want functions to be \lstinline|Proper|, which means that they respect equality. For setoids with apartness we need a stronger property, \emph{strong extensionality}.
\begin{lstlisting}
Class StrongSetoid_Morphism {A B} {Ae : Equiv A} {Aap : Apart A}
    {Be: Equiv B} {Bap : Apart B} (f : A → B):= { 
  strong_setoidmor_a: StrongSetoid A;
  strong_setoidmor_b: StrongSetoid B;
  strong_extensionality : ∀ x y, f x ⪥ f y → x ⪥ y }.
\end{lstlisting}
We prove that for each \lstinline|StrongSetoid_Morphism f| we have \lstinline|Proper ((=) ==> (=)) f|. The only structures for which we actually need apartness are implementations of the real numbers, hence we only base the \lstinline|Field| class on top of a \lstinline|StrongSetoid| instead of the complete algebraic hierarchy. Our class for fields is as follows. (The \lstinline|PropHolds| class is explained in the next subsection.)
\begin{lstlisting}
Class Recip A `{Apart A} `{Zero A} := recip: { x : A | x ⪥  0 } → A.
Notation "// x" := (recip x).
Notation "(//)" := recip (only parsing).

Class Field A {Ae Aplus Amult Azero Aone Aneg} {Aap : Apart A} {Arecip : Recip A} : Prop := {
  field_ring :> @Ring A Ae Aplus Amult Azero Aone Aneg;
  field_strongsetoid :> StrongSetoid A;
  field_plus_ext :> StrongSetoid_BinaryMorphism (+);
  field_mult_ext :> StrongSetoid_BinaryMorphism (.*.);
  field_nontrivial :> PropHolds (1 ⪥ 0);
  recip_proper :> Setoid_Morphism (//);
  recip_inverse : ∀ x, proj1_sig x // x = 1 }.
\end{lstlisting}
We do not include strong extensionality of the inverse and the reciprocal because these properties can be derived.

\begin{figure}[hpt!]
	\subfloat[The algebraic hierarchy]{\includegraphics[width=0.4\textwidth]{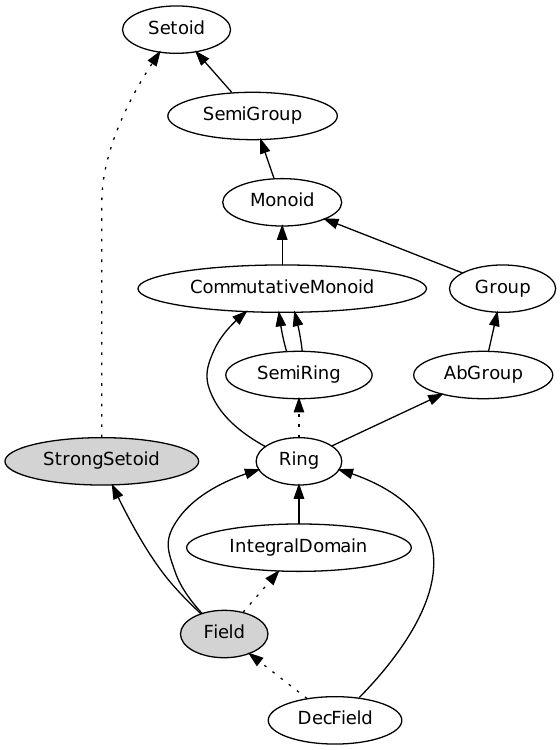}}
	\subfloat[The order hierarchy]{\includegraphics[width=0.6\textwidth]{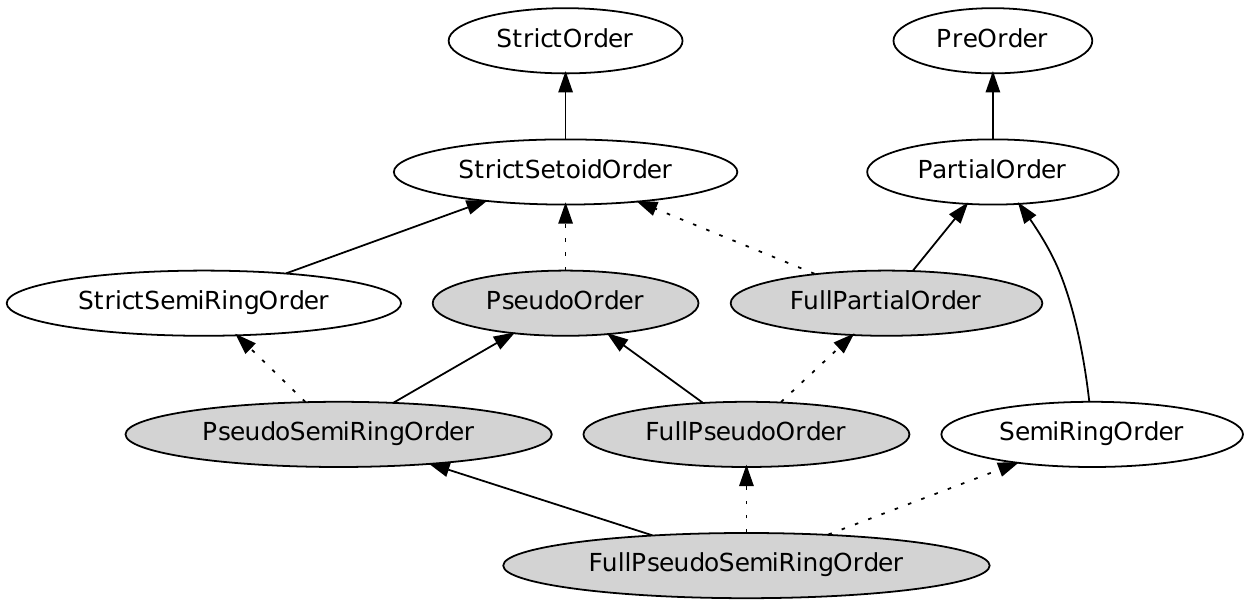}}
	\caption{The algebraic and order hierarchy. Dotted lines denote derived inheritance, filled nodes denote presence of apartness.}
	\label{figure:hierarchy}
\end{figure}

For convenience, we define an additional class \lstinline|DecField| for fields with decidable equality and whose reciprocal function is total. This class integrates nicely with \Coq's rational numbers \lstinline|Q| and \lstinline|bigQ|, and the \lstinline|field| tactic to solve field equations. This total reciprocal function should satisfy $/0 = 0$, so properties as $f(/x) = /(f x)$, $/(/x) = x$ and $/x * /y = /(x * y)$ hold without any additional premises. We proved that a \lstinline|DecField| is also an instance of our \lstinline|Field| class. A diagram of our complete algebraic hierarchy is displayed in Figure~\ref{figure:hierarchy}.

\subsection{Order theory}
Existing \Coq{} libraries for ordered algebraic structures turn out to be too limited to abstract from $\N$, $\Z$, $\Q$ and $\R$ and their various implementations. The formalization of ordered fields in the CoRN library~\cite{C-corn} restricts to a very specific part of the algebraic hierarchy (namely fields). Letouzey's \Name{Numbers} library, which is included in recent versions of \Coq{} trunk, only considers $\N$ and $\Z$. The \Ssreflect{} library presently restricts to decidable structures with Leibniz equality. Moreover, even mathematically, the most convenient  abstraction is not entirely clear. Le\v{s}nik~\cite{Lesnik:Phd} provides a smooth order theoretic characterization of these structures as so-called \emph{streaks}. We, however, prefer our theory below as it avoids partial functions.

In this work we present a library that captures the notion of order on a variety of structures, including structures with undecidable equality. One of the building blocks of our hierarchy is a pseudo order~\cite{Heyting:1956}, which is the constructive variant of a total order.
\begin{lstlisting}
Class PseudoOrder `{Ae : Equiv A} `{Aap : Apart A} (Alt : Lt A) : Prop := {
  pseudo_order_setoid : StrongSetoid A; 
  pseudo_order_asym : ∀ x y, ¬(x < y ∧ y < x);
  pseudo_order_cotrans :> CoTransitive (<);
  apart_iff_total_lt : ∀ x y, x ⪥ y ↔ x < y ∨ y < x }.
\end{lstlisting}
In case equality is decidable, this interface is rather awkward to work with. Therefore we present ways to go back and forth between the usual classical notions and their constructive variants. For example, we use the type class machinery to infer the classical trichotomy property in case apartness is just the negation of equality.
\begin{lstlisting}
Instance lt_trichotomy `{PseudoOrder A} `{!TrivialApart A} `{∀ x y, Decision (x = y)} : Trichotomy (<).
\end{lstlisting}
Also, we can go the other way around. If we have a \lstinline|StrictSetoidOrder| (an ordinary strict order built upon a setoid) satisfying the trichotomy property, we obtain a pseudo order.
\begin{lstlisting}
Lemma dec_strict_pseudo_order `{StrictSetoidOrder A} `{Apart A} 
   `{!TrivialApart A} `{∀ x y, Decision (x = y)} `{!Trichotomy (<)} : PseudoOrder (<).
\end{lstlisting}
In order to avoid loops, we define the above as an ordinary \lstinline|Lemma| instead of an \lstinline|Instance|. Next, one could extend a pseudo order to the standard notion of a (pseudo) ring order.
\begin{lstlisting}
Class PseudoRingOrder `{Equiv A} `{Apart A} `{Plus A}
    `{Mult A} `{Zero A} `{One A} `{Negate A} (Alt : Lt A) := { 
  pseudo_ringorder_spo :> PseudoOrder Alt;
  pseudo_ringorder_ring : Ring A;
  pseudo_ringorder_mult_ext :> StrongSetoid_BinaryMorphism (.*.);
  pseudo_ringorder_plus :> ∀ z, StrictlyOrderPreserving (z +);
  pseudo_ringorder_mult : ∀ x y, 0 < x → 0 < y → 0 < x * y }.
\end{lstlisting}
However, we wish to use our library on ordered structures for implementations of the natural numbers as well. Since the natural numbers do not form a ring, but merely a semiring, we strengthen the above class with a partial subtraction function (living in \Prop{}, because we never use it for computations) and require addition to be order reflecting. We call this, apparently new notion, a \lstinline|PseudoSemiRingOrder|. 
\begin{lstlisting}
Class PseudoSemiRingOrder `{Equiv A} `{Apart A} `{Plus A} 
    `{Mult A} `{Zero A} `{One A} (Alt : Lt A) := { 
  pseudo_srorder_strict :> PseudoOrder Alt;
  pseudo_srorder_semiring : SemiRing A;
  pseudo_srorder_partial_minus : ∀ x y, ¬y < x → ∃ z, y = x + z;
  pseudo_srorder_plus :> ∀ z, StrictOrderEmbedding (z +);
  pseudo_srorder_mult_ext :> StrongSetoid_BinaryMorphism (.*.);
  pseudo_srorder_pos_mult_compat : ∀ x y,
    PropHolds (0 < x) → PropHolds (0 < y) → PropHolds (0 < x * y) }.
\end{lstlisting}
Instead of including the \lstinline|PseudoRingOrder| class in our development, we include a lemma to construct a \lstinline|PseudoSemiRingOrder| from a ring satisfying the \lstinline|PseudoRingOrder| axioms. 

Given a pseudo (semiring) order, one could define the non-strict order \lstinline|x ≤ y| in terms of the strict order, namely as \lstinline|¬y < x|. However, this is quite inconvenient in practice, because we also want to talk about a priori different non-strict orders such as those defined in the standard library. Hence we introduce the following class.
\begin{lstlisting}
Class FullPseudoSemiRingOrder `{Equiv A} `{Apart A} `{Plus A} 
    `{Mult A} `{Zero A} `{One A} (Ale : Le A) (Alt : Lt A) := { 
  full_pseudo_srorder_pso :> PseudoSemiRingOrder Alt;
  full_pseudo_srorder_le_iff_not_lt_flip : ∀ x y, x ≤ y ↔ ¬y < x }.
\end{lstlisting}
A diagram of our complete order hierarchy is displayed in Figure \ref{figure:hierarchy}. 

Our theory on abstract orders avoids duplication of theorems and proofs. For example, the following lemmas apply to $\N$, $\Z$, $\Q$ and the dyadics, because all of these structures form a \lstinline|FullPseudoSemiRingOrder|.
\begin{lstlisting}
Lemma plus_compat x1 y1 x2 y2 : x1 ≤ y1 → x2 ≤ y2 → x1 + x2 ≤ y1 + y2.
Lemma lt_1_2 : 1 < 2.
Lemma square_nonneg x : 0 ≤ x * x.
\end{lstlisting}
To allow us to refer by canonical names to common properties, we introduce classes like:
\begin{lstlisting}
Class OrderPreserving {A B} {Ae : Equiv A} {Ale : Le A} {Be : Equiv B} {Ble : Le B} (f : A → B) := {
  order_preserving_morphism :> Order_Morphism;
  order_preserving : ∀ x y, x ≤ y → f x ≤ f y }.
Class OrderReflecting {A B} {Ae : Equiv A} {Ale : Le A} {Be : Equiv B} {Ble : Le B} (f : A → B) := {
  order_preserving_back_morphism :> Order_Morphism;
  order_preserving_back : ∀ x y, f x ≤ f y → x ≤ y }.
\end{lstlisting}
Here, an \lstinline|Order_Morphism| is just the factoring out of the common parts of both classes; namely that \lstinline|f| and \lstinline|≤| respect equality. For the case of multiplication these properties have additional premises, for example:
\begin{lstlisting}
Global Instance: ∀ (z : A), PropHolds (0 < z) → OrderPreserving (z *.).
\end{lstlisting}
We introduce the \lstinline|PropHolds| class to let the type class machinery prove these properties automatically. For example consider:
\begin{lstlisting}
Lemma example (n : N) (x y : A) : x ≤ y → (2 ^ n + 2) * x ≤ (2 ^ n + 2) * y.
Proof. intros. now apply (order_preserving (2 ^ n + 2 .*)). Qed.
\end{lstlisting}
In order to use \lstinline|order_preserving|, we need a proof of \lstinline|PropHolds (0 < 2 ^ n + 2)|. Type class resolution is able to prove this in a fully automated way because we have the following instances:
\begin{lstlisting}
Instance: PropHolds (0 < 2);
Instance: ∀ x y : A, PropHolds (0 < x) → PropHolds (0 < y) → PropHolds (0 < x + y)
Instance: ∀ (n : N) (x : A), PropHolds (0 < x) → PropHolds (0 < x ^ n)
\end{lstlisting}
This example shows that type class search is in fact very similar to proof search by the \lstinline|auto| tactic, but there is no need to call a tactic by hand.

For arbitrary instances of $\N$, $\Z$, $\Q$ it is easy to define an order satisfying these interfaces:
\begin{lstlisting}
Instance nat_le `{Naturals N} : Le N | 10 :=  λ x y, ∃ z, y = x + z.
Instance nat_lt `{Naturals N} : Lt N | 10 :=  λ x y, x ≤ y ∧ x ≠ y.
\end{lstlisting}
However, often we encounter an a priori different order on a structure, most likely an order defined in \Coq's standard library (like \lstinline|Nle| and \lstinline|Nlt| on \lstinline|N|). Therefore we prove that a \lstinline|FullPseudoSemiRingOrder| uniquely specifies the order on $\N$, $\Z$ and $\Q$. For example:
\begin{lstlisting}
Context `{Naturals N} `{Naturals N2} {f : N → N2} `{!SemiRing_Morphism f}
  `{Apart N} `{!TrivialApart N} `{!FullPseudoSemiRingOrder (A:=N) Nle Nlt} 
  `{Apart N2} `{!TrivialApart N2} `{!FullPseudoSemiRingOrder (A:=N2) N2le N2lt}.
Global Instance: OrderEmbedding f.
\end{lstlisting}
Unfortunately \Coq{} has no support to have an argument be `inferred if possible, generalized otherwise'; see~\cite{math-classes}. When declaring an argument of \lstinline|FullPseudoSemiRingOrder|, one is often in a context where most of its components are already available. Here, only the additional arguments \lstinline|Le|, \lstinline|Lt| and \lstinline|Apart| have to be introduced. The current workaround in these cases (as shown above) involves providing names for components that are then never referred to, which is a bit awkward. In the above it would much nicer to write:
\begin{lstlisting}
Context `{Naturals N} `{Naturals N2} {f : N → N2} `{!SemiRing_Morphism f}
  `{!TrivialApart N} `{!FullPseudoSemiRingOrder N} `{!TrivialApart N2} `{!FullPseudoSemiRingOrder N2}.
Global Instance: OrderEmbedding f.
\end{lstlisting}

\subsection{Basic operations}
The operation \lstinline|nat_pow| is most commonly, but inefficiently, defined as repeated multiplication and the operation \lstinline|shiftl| is defined as repeated duplication. Instead we specify the desired behavior of these operations. This approach allows for different implementations for different number representations and avoids definitions and proofs  becoming implementation dependent.

We introduce interfaces that specify the behavior of the operations \lstinline|abs|, \mbox{\lstinline|shiftl|,} \lstinline|nat_pow| and \lstinline|int_pow|. Again there are various ways of specifying these interfaces: with $\Sigma$-types, bundled or unbundled. In general, \mbox{$\Sigma$-types} are convenient for functions whose specification is easy, for example:
\begin{lstlisting}
Class Abs A `{Equiv A} `{Le A} `{Zero A} `{Negate A} 
	:= abs_sig: ∀ (x : A), { y : A | (0 ≤ x → y = x) ∧ (x ≤ 0 → y = -x)}.
Definition abs `{Abs A} := λ x : A, proj1_sig (abs_sig x).
\end{lstlisting} 
However, for more complex operations, such as \lstinline|shiftl|, we follow the unbundled approach by Spitters and van der Weegen~\cite{math-classes}.
\begin{lstlisting}
Class ShiftL A B := shiftl: A → B → A.
Infix "≪" := shiftl (at level 33, left associativity).
Class ShiftLSpec A B (sl : ShiftL A B) `{Equiv A} `{Equiv B} `{One A} 
	`{Plus A} `{Mult A} `{Zero B} `{One B} `{Plus B} := {
  shiftl_proper : Proper ((=) ==> (=) ==> (=)) (≪);
  shiftl_0 :> RightIdentity (≪) 0;
  shiftl_S : ∀ x n, x ≪ (1 + n) = 2 * x ≪ n }.
\end{lstlisting}
We do not specify \lstinline|shiftl| as \lstinline|shiftl x n = x * 2 ^ n| since on the dyadics we cannot take a negative power while we can shift by a negative integer. Since theory on shifting with exponents in $\N$ and $\Z$ is similar we want to avoid duplication of theorems and proofs. To this end we introduce a class describing the bi-induction principle.
\begin{lstlisting}
Class Biinduction A `{Equiv A} `{Zero A} `{One A} `{Plus A} : Prop 
  := biinduction (P: A → Prop) `{!Proper ((=) ==> iff) P} : P 0 → (∀n, P n ↔ P (1 + n)) → ∀ n, P n.
\end{lstlisting}
Since this class is inhabited by any integer and natural implementation we can parametrize theory on \lstinline|shiftl| as follows.
\begin{lstlisting}
Context `{SemiRing A} `{!LeftCancellation (.*.) (2:A)} `{SemiRing B} `{!Biinduction B} `{!ShiftLSpec A B sl}.
Lemma shiftl_base_plus x y n : (x + y) ≪ n  = x ≪ n + y ≪ n.
Global Instance shiftl_inj: ∀ n, Injective ($≪$n).
\end{lstlisting}
	
\subsection{Decision procedures}
\label{section:decision}
The \lstinline|Decision| type class by Spitters and van der Weegen collects decidable propositions~\cite{math-classes}.
\begin{lstlisting}
Class Decision P := decide: sumbool P (¬P).
\end{lstlisting}
Using this type class we can declare a argument \lstinline|`{∀ x y, Decision (x = y)}| to describe a decider for $=$ and say \lstinline|decide (x = y)| to decide whether \lstinline|x = y| or not. This type class allows us to easily compose deciders, for example:
\begin{lstlisting}
Instance prod_dec `(A_dec : ∀ x y : A, Decision (x = y))
  `(B_dec : ∀ x y : B, Decision (x = y)) : ∀ x y : A * B, Decision (x = y).
\end{lstlisting}
We have to be careful however. Consider the definition of the order on the dyadics.
\begin{lstlisting}
Global Instance dy_le: Le Dyadic := λ x y : Dyadic, 
  ZtoQ (mant x) * 2 ^ (expo x) ≤ ZtoQ (mant y) * 2 ^ (expo y)
Global Instance dy_le_dec: ∀ (x y : Dyadic), Decision (x ≤ y).
\end{lstlisting}
Now, \lstinline|decide (x ≤ y)| for \lstinline|x| and \lstinline|y| of type \lstinline|Dyadic| is actually \lstinline|@decide (x ≤ y)$\ $(dy_le_dec x y)|. This shows that the proposition \lstinline|x ≤ y| is just a phantom argument used for instance search only, whereas \lstinline|dy_le_dec| is the decision procedure doing the actual work. Due to eager evaluation of \Coq's virtual machine, the term \lstinline|decide (x ≤ y)| is expanded to
\begin{center}
\lstinline|@decide (ZtoQ (mant x) * 2 ^ (expo x) ≤ ZtoQ (mant y) * 2 ^ (expo y))$\ $(dy_le_dec x y)|,
\end{center}
resulting in the phantom argument being evaluated first.
In many cases evaluation of such a phantom argument is cheap, but here it involves an expensive conversion of \lstinline|x| and \lstinline|y| to \lstinline|Q|. We avoid evaluation of this phantom argument by wrapping it under a λ-abstraction.
\begin{lstlisting}
Definition decide_rel `(R : relation A) {dec : ∀ x y, Decision (R x y)} 
  (x y : A) : Decision (R x y) := dec x y.
\end{lstlisting}
Now, if we write \lstinline|decide_rel (≤) x y|, it expands to 
\begin{center}
\lstinline|@decide_rel (λ x y, ZtoQ (mant x) * 2 ^ (expo x) ≤ ZtoQ (mant y) * 2 ^ (expo y))$\ $x y  dy_le_dec|,
\end{center}
where the definition of inequality is safely hidden under a λ-abstraction.

This problem would not appear if \Coq{}'s virtual machine would evaluate propositions lazily, as the phantom argument is just a proposition. Unfortunately, lazy evaluation of propositions is not supported by its current implementation.

\subsection{Explicit type casts}
The \lstinline|Cast| type class collects (explicit) type casts.
\begin{lstlisting}
Class Cast A B := cast: A → B.
Implicit Arguments cast [[Cast]].
Notation "' x" := (cast _ _ x) (at level 20).
Instance: Params (@cast) 3.
\end{lstlisting}
This definition allows us to refer to a cast from \lstinline|x : A| to \lstinline|B| by using an apostrophe, or writing \lstinline|cast A B x|. An example of an instance of this class is:
\begin{lstlisting}
Instance NonNeg_inject: Cast (A⁺) A := @proj1_sig A _.
\end{lstlisting}
Here, $\nonneg A$ is a $\Sigma$-type describing the non-negative cone of an ordered ring $A$. Contrary to \Coq's built-in coercion mechanism, our type casts are explicit instead of implicit and type classes are used to register them. Our approach has some advantages:
\begin{enumerate}[(1)]
\item By using type classes to register casts, we are allowed to parametrize classes with casts. An example is the \lstinline|AppRationals| class, as defined in Section \ref{section:approx_rationals}.
\item Implicit coercions often introduce ambiguity. Since our approach allows us to refer to casts by a (canonical) name, e.g.\ \lstinline|cast B C (cast A B x)|, we can avoid this ambiguity.
\item Casts can be put in partially applied position, e.g.\ \lstinline|order_preserving (cast Z Q)|.
\end{enumerate}

\Coq's coercion mechanism does not allow us to define a coercion from $\nonneg A$ to $A$ nor a coercion from a ring to its polynomial ring. More generally, it does not allow most forms of parametrized coercions nor non-uniform coercions.
An implementation that allows parametrized coercions like \lstinline|NonNeg_inject| has to avoid an infinite loop: to naively type check \lstinline|x : A|, one has to type check \lstinline|x : A⁺|, \lstinline|x : (A⁺)⁺|, \ldots{} We suffer from such loops if we compose our \lstinline|Cast| classes automatically as well. Hence we refrain from adding:
\begin{lstlisting}
Instance cast_comp_base `{f : Cast A B} : ComposedCast A B := f.
Instance cast_comp_step `{f : Cast B C} `{g : ComposedCast A B} : ComposedCast A C := λ x, f (g x).
\end{lstlisting}
Matita~\cite{asperti2007user} allows parametrized coercions and avoids the loop by not applying coercions recursively, but instead building a well-chosen set of set of composite coercions~\cite{tassi-phd}. Non-uniform coercions~\cite{non-unif-coerc} are available in Matita. They are implemented using unification hints, a feature similar to type classes.

\section{The real numbers}\label{section:reals}
\label{section:approx_rationals}
To make our implementation of the reals independent of the underlying dense set, we provide an abstract specification of \emph{approximate rationals} inspired by the notion of \emph{approximate fields} --- a field with approximate operations --- which is used in Bauer and Kavler's RZ implementation of the exact reals~\cite{bauer2008implementing}; see also~\cite{bauer2009dedekind}. In particular, we provide an implementation of this interface by dyadics based on \Coq's machine integers.

Our interface for approximate rationals describes an ordered ring containing \lstinline|Z| that is dense in \lstinline|Q|. Here \lstinline|Z| are the binary integers from \Coq's standard library, and \lstinline|Q| are the rationals based on these binary integers. We do not parametrize by arbitrary integer and rational implementations because they are hardly used for computation. For efficient computation we include the operations: approximate division, normalization, an embedding of \lstinline|Z|, absolute value, power by \lstinline|N|, shift by \lstinline|Z|, and decision procedures for equality and order.
\begin{lstlisting}
Class AppDiv AQ := app_div: AQ → AQ → Z → AQ.
Class AppApprox AQ := app_approx: AQ → Z → AQ.
Class AppRationals AQ {AQe AQplus AQmult AQzero AQone AQneg} `{Apart AQ} `{Le AQ} `{Lt AQ}
     {AQtoQ : Cast AQ Q_as_MetricSpace} `{!AppInverse AQtoQ} {ZtoAQ : Cast Z AQ} 
     `{!AppDiv AQ} `{!AppApprox AQ} `{!Abs AQ} `{!Pow AQ N} `{!ShiftL AQ Z} 
     `{∀ x y : AQ, Decision (x = y)} `{∀ x y : AQ, Decision (x ≤ y)} : Prop := {
  aq_ring :> @Ring AQ AQe AQplus AQmult AQzero AQone AQneg;
  aq_trivial_apart :> TrivialApart AQ;
  aq_order_embed :> OrderEmbedding AQtoQ;
  aq_strict_order_embed :> StrictOrderEmbedding AQtoQ;
  aq_ring_morphism :> SemiRing_Morphism AQtoQ;
  aq_dense_embedding :> DenseEmbedding AQtoQ;
  aq_div : ∀ x y k, ball (2 ^ k) ('app_div x y k) ('x / 'y);
  aq_compress : ∀ x k, ball (2 ^ k) ('app_approx x k) ('x);
  aq_shift :> ShiftLSpec AQ Z (≪);
  aq_nat_pow :> NatPowSpec AQ N (^);
  aq_ints_mor :> SemiRing_Morphism ZtoAQ }.
\end{lstlisting}
We define the real numbers as the completion of the approximate rationals. To create functions on the real numbers, we use the monadic operations \bind{} or \map{}. This approach is convenient because equality and inequality are decidable on the approximate rationals, whereas it is not on the real numbers. For binary functions, e.g. addition and multiplication, we use the \lstinline|map2| function, as described in~\cite{OConnor:mscs}.

O'Connor~\cite{OConnor:mscs} keeps the size of the rational numbers small to avoid efficiency problems. He introduces a function \mbox{\lstinline|approx x ε|} that yields the `simplest' rational number between \lstinline|x - ε| and \lstinline|x + ε|. We modify the \lstinline|approx| function slightly: \lstinline|app_approx x k| yields an arbitrary element between \lstinline|x - $2^k$| and \mbox{\lstinline|x + $\ 2^k$|.}
Using this function we define the compress operation on the real numbers: \mbox{\lstinline|compress := bind (λ x ε, app_approx x (Qdlog2 ε))|} such that \mbox{\lstinline|compress x = x|}.

In Section~\ref{section:series} we will explain our choice of using a power of 2 to specify the precision of \lstinline|app_div| and \lstinline|app_approx|. 

\subsection{Order and apartness}
\label{section:reals_order}
Following~\cite{Bishop/Bridges:1985,oconnor-thesis}, we define non-negativity and the order on the real numbers as follows.
\begin{flalign*}
	\NonNeg\ x &:= ∀ ε : \pos\Q, -ε ≤ x\ ε \\
	x ≤ y &:= \NonNeg\ (y - x)
\end{flalign*}
Bishop and Bridges~\cite{Bishop/Bridges:1985} define positivity as the dual of non-negativity: \mbox{$∃ ε : \pos\Q,\ ε < x\ ε$}. O'Connor~\cite{oconnor-thesis} defines positivity and the strict order differently so as to avoid a potentially expensive computation, namely $x\ ε - ε$, to obtain a witness between $0$ and $x$.
\begin{flalign*}
	\Pos\ x &:= \{ ε : \pos\Q \separator ε ≤ x \} \\
	x <_\T y &:= \Pos\ (y - x)
\end{flalign*}
We use the $\T$ subscript to emphasize that the relation lives in \Type. Next, we define $x \apart_\T y := x <_\T y ∨ y <_\T x$. Extraction of a witness $ε \in (0, x]$ from $\Pos\ x$ allows us to define the reciprocal function of type $∀ x : \R, 0 \apart_\T x → \R$. 

In order to use our type class based hierarchy we need a strict order and apartness relation in \Prop. We need this restriction because \Coq's present implementation of setoid rewriting does not allow rewriting in \Type-based relations (see Section~\ref{section:apartness}). Our definition is similar to Bishop and Bridges' definition of positivity, but uses shifts instead.
\begin{flalign*}
	x < y &:= ∃ n : \N,\ 1 ≪ -n < (y - x)\ (1 ≪ (-n - 1))\\
	x \apart y &:= x < y ∨ y < x
\end{flalign*}
Using constructive indefinite description (see Section~\ref{section:indefinitive}), it is an easy job to prove that we indeed have $x < y ↔ x <_\T y$ and $x \apart y ↔ x \apart_\T y$. Similar to O'Connor~\cite{oconnor-thesis}, we implement a tactic that automatically proves strict inequalities. The tactic terminates iff the inequality holds and is similar to our use of linear search to obtain $x <_\T y$ from $x < y$.

\subsection{Implementation using the dyadics}
\label{section:dyadics}
The dyadic rationals are numbers of the shape $n * 2 ^ e$ for $n,e \in \Z$. In order to remain independent of a specific implementation of integers, we have defined most of the operations for arbitrary integer implementations.
Given such an implementation \lstinline|Int| it is straightforward to define the ring operations.
\begin{lstlisting}[mathescape=false,breaklines=false]
Notation "x↾p" := (exist _ x p) (at level 20).
Record Dyadic := dyadic { mant : Int; expo : Int }.
Infix "▼" := dyadic (at level 80).
Global Instance dy_inject: Cast Int Dyadic := λ x, x ▼ 0.
Global Instance dy_negate: Negate Dyadic := λ x, -mant x ▼ expo x.
Global Instance dy_mult: Mult Dyadic := λ x y, mant x * mant y ▼ expo x + expo y.
Global Instance dy_0: Zero Dyadic := cast Int Dyadic 0.
Global Instance dy_1: One Dyadic := cast Int Dyadic 1.
Global Program Instance dy_plus: Plus Dyadic := λ x y, 
  if decide_rel (≤) (expo x) (expo y)
  then mant x + mant y ≪ (expo y - expo x)↾_ ▼ min (expo x) (expo y)
  else mant x ≪ (expo x - expo y)↾_ + mant y ▼ min (expo x) (expo y).
\end{lstlisting}
In this code \lstinline|(≪)| has type \lstinline|Int → Int⁺ → Int|, where \lstinline|Int⁺| is a $\Sigma$-type describing the non-negative cone of \lstinline|Int|. Therefore, in the definition of \lstinline|dy_plus| we have to equip \lstinline|expo y - expo x| with a proof that it is in fact non-negative.

The operation of approximate division is not implemented in an abstract way as we have not developed a type class and theory for right shifts yet. For our implementation using \Coq's machine integers \lstinline|bigZ|, we defined approximate division concretely using the shift right function from the standard library.

\subsection{Implementation using the rationals}
\label{section:fast_rationals}
Our development contains additional implementations of the \lstinline|AppRationals| class using \Coq's old rational numbers \lstinline|Q| and the new rational numbers \lstinline|bigQ| (which are built from the machine integers \lstinline|bigZ|). Although creating these implementations is uninteresting from a performance point of view, it confirms that it is trivial to change the underlying dense set from which our real numbers are built. 

To implement the \lstinline|app_approx| function in an efficient manner, we use shifts on the underlying integers. Furthermore, to keep the size of the results of the division operation small, we incorporate the \lstinline|app_approx| function.
\begin{lstlisting}
Instance bigQ_div: AppDiv bigQ := λ x y, app_approx (x / y).
\end{lstlisting}

\subsection{Power series}
\label{section:series}
Elementary transcendental functions as \exp, \sin, \ln{} and \atan{} can be defined by their power series. If the coefficients of a power series are alternating, decreasing and have limit 0, then we obtain a fast converging sequence with an easy termination proof. For $-1 ≤ x ≤ 0$,
\[
	\exp\ x = \sum_{i=0}^\infty \frac {x^i} {i!}
\]
is of this form. To approximate $\exp\ x$ with error $ε$ we take the partial sum until $\frac{x^i}{i!} ≤ ε$.
In order to implement this efficiently we use a stream representing the series and define a function that sums the required number of elements. For example, the series \lstinline|1, a, a$^2$, a$^3$, ...| is defined by the following stream.
\begin{lstlisting}
CoFixpoint powers_help (c : A) : Stream A := Cons c (powers_help (c * a)).
Definition powers : Stream A := powers_help 1.
\end{lstlisting}
Streams in \Coq, like lists in \Haskell, are lazy. So, in the example the multiplications are accumulated.

Since \Coq{} only allows structural recursion (and guarded co-recursion) it requires some work to convince \Coq{} that our algorithm terminates. Intuitively, one would describe the limit as an upperbound of the required number of elements using the \lstinline|Exists| predicate.
\begin{lstlisting}
Inductive Exists A (P : Stream A → Prop) (x : Stream) : Prop :=
  | Here : P x → Exists P x
  | Further : Exists P (tl x) → Exists P x.
\end{lstlisting}
This approach leads to performance problems. The upperbound, encoded in unary format, may become very large while generally only a few terms are necessary. Due to \lstinline|vm_compute|'s eager evaluation scheme, this unary number will be computed before summing the series. Instead O'Connor~\cite{oconnor-thesis} uses \lstinline|LazyExists|.
\begin{lstlisting}
Inductive LazyExists A (P : Stream A → Prop) (x : Stream A) : Prop :=
  | LazyHere : P x → LazyExists P x
  | LazyFurther : (unit → LazyExists P (tl x)) → LazyExists P x.
\end{lstlisting}
Unfortunately, our experiments showed that the above still yields too much overhead due unnecessary to reduction of proofs. To remedy this issue we introduce the following function where
\lstinline|Str_nth_tl $n$ $s$| takes the $n$-th tail of the stream $s$.
\begin{lstlisting}
Fixpoint LazyExists_inc `{P : Stream A → Prop}
    (n : nat) s : LazyExists P (Str_nth_tl n s) → LazyExists P s :=
  match n return LazyExists P (Str_nth_tl n s) → LazyExists P s with
  | O => λ x, x
  | S n => λ ex, LazyFurther (λ _, LazyExists_inc n (tl s) ex)
  end. 
\end{lstlisting}
This function adds \lstinline|n| additional \lstinline|LazyFurther| constructors. When instantiated with a big enough \lstinline|n|, computation will suffer from the implementation limits of \Coq{} (e.g.\ a stack overflow) or runs out of memory, before it ever refers to the actual proof. Using \lstinline|LazyExists_inc| we are able to compute on average twice the amount of decimals as we did before on examples such as the ones in Table~\ref{table:coq}.

O'Connor's \lstinline|InfiniteAlternatingSum $s$| returns the real number represented by the infinite alternating sum over $s$, where the stream $s$ is decreasing, non-negative and has limit 0. We extend this in two ways. First, we generalize various notions to abstract structures. Secondly, as we do not have exact division on approximate rationals, we extend the algorithm to work with approximate division. The latter requires changing \lstinline|InfiniteAlternatingSum $s$| to \lstinline|InfiniteAlternatingSum $n\ d$| which computes the infinite alternating sum of the stream $λ i,\frac {n_i}{d_i}$. This allows us to postpone divisions. Also, we have to determine both the length of the partial sum and the required precision of the divisions. To do so we find a $k$ such that:
\begin{equation}
	\ball {\frac{ε}{2}} {(\appdiv\ n_k\ d_k\ (\log \frac{ε}{2k}) + \frac{ε}{2k})} 0.\label{equation:series_ball}
\end{equation}
Now $k$ is the length of the partial sum, and $\frac{ε}{2k}$ is the required precision of division. Using O'Connor's results we have verified that these values are correct and such a $k$ indeed exists for a decreasing, non-negative stream with limit 0.

As noted in Section~\ref{section:approx_rationals}, we have specified the precision of division in powers of 2 instead of using a rational value. This allows us to replace (\ref{equation:series_ball}) with:
\[
	\ball {\frac{ε}{2}} {(\appdiv\ n_k\ d_k\ (\log\ ε - (k + 1)) + 1 ≪ (\log\ ε - (k + 1)))} 0.
\]
Here $k$ is the length of the partial sum, and $2^l$, where $l = \log\ ε - (k + 1)$, is the required precision of division. This variant can be implemented without any arithmetic on the rationals and is thus much more efficient.

This method gives us a fast way to compute the infinite alternating sum, in practice, only a few extra terms have to be computed and due to the approximate division the auxiliary results are kept as small as possible.

Similarly, using this method to compute infinite alternating sums, we use the following series to implement $\atan\ x$ and $\sin\ x$ for $x \in [-1, 1]$.
\begin{flalign*}
	\atan\ x &= \sum_{i=0}^\infty (-1)^i * \frac{x^{2i + 1}} {(2i + 1)!} \\
	\sin\ x &= \sum_{i=0}^\infty (-1)^i * \frac{x^{2i + 1}} {2i + 1}
\end{flalign*}
We extend these functions to their complete domain by repeatedly applying the following formulas~\cite{oconnor-thesis}.
\begin{flalign}
	\exp\ x &= (\exp\ (x ≪ 1))^2 \label{equation:exp_squaring} \\
	\exp\ x &= \frac{1}{\exp\ (-x)} \\
	\sin\ x &= 3 * \sin\ \frac{x}{3} - 4 * \Big(\sin\ \frac{x}{3}\Big)^3 \label{equation:sin_reduce}\\
	\atan\ x &= -\atan\ (-x) \\
	\atan\ x &= \frac{\pi}{2} - \atan\ \frac{1}{x} \quad\textnormal{for $0 < x$} \label{equation:atan_reduce1}\\
	\atan\ x &= \frac{\pi}{4} - \atan\ \Big(\frac{x - 1}{x + 1}\Big) \quad\textnormal{for $0 < x$} \label{equation:atan_reduce2}
\end{flalign}
Since we do not have exact division on the approximate rationals, we parameterize infinite sums by two streams in Equation~\ref{equation:sin_reduce}, \ref{equation:atan_reduce1} and~\ref{equation:atan_reduce2}. 

The series described in this section converge faster for arguments closer to 0. We use Equation \ref{equation:exp_squaring} and \ref{equation:sin_reduce} repeatedly to reduce the input to a value \mbox{$|x| \in [0, 2^k)$}. For $50 ≤ k$, this yields nearly always major performance improvements, for higher precisions setting it to $75 ≤ k$ yields even better results.  Unfortunately, we are unaware of a similar trick for \atan. We define $\pi$ in terms of \atan{} using the following Machin-like formula.
\[
	\pi := 176 * \atan \frac{1}{57} + 28 * \atan \frac{1}{239} 
		- 48 * \atan \frac{1}{682} + 96 * \atan \frac{1}{12943}
\]
Again, here we notice the purpose of parameterizing infinite sums by two streams. We define \cos{} in terms of \sin{}.
\[
	\cos\ x = 1 - 2 * \Big(\sin\ \frac{x}{2}\Big)^2 \\
\]
O'Connor~\cite{OConnor:mscs, oconnor-thesis} subtracts multiples of $2\pi$ to reduce the arguments of \sin{} and \cos. In our tests this did not lead to performance improvements because our implementation of $\pi$ turned out to be slower than the performed range reductions.

\subsection{Square root}
\label{section:Wolfram}
We use Wolfram's algorithm~\cite[p.913]{wolfram2002new} for computing the square root. Its complexity is linear, in fact it provides a new binary digit in each step.
\begin{lstlisting}
Context `(Pa : 1 ≤ a ≤ 4).
Fixpoint AQroot_loop (n : nat) : AQ * AQ :=
  match n with
  | O => (a, 0)
  | S n =>
     let (r, s) := AQroot_loop n in
     if decide_rel (≤) (s + 1) r
     then ((r - (s + 1)) ≪ (2:Z), (s + 2) ≪ (1:Z))
     else (r ≪ (2:Z), s ≪ (1:Z))
  end.
\end{lstlisting}
We write $(r_n,s_n)$ for the $n$-th pair of approximations. By  induction we obtain:
\begin{flalign}
 s_n^2+4r_n &= 4^{n+1}a\label{equation:wolfram1}\\
 r_n &≤ 2 s_n +4\\
 2^m s_n ≤ s_{n+m} &≤ 2^m(s_n +4)-4\label{equation:wolfram3}\\
 r_n &≤ 2 ^ {3 + n}\label{equation:wolfram4}
\end{flalign}
By \ref{equation:wolfram1}, $(2^{-(n+1)} s_n)^2 + 2^{-2n}r_n=a$. By \ref{equation:wolfram4}, $2^{-2n}r_n$ converges to 0 as $n$ tends to $\infty$. Therefore, by \ref{equation:wolfram3}, $2^{-(n+1)} s_n$ is a Cauchy sequence which moreover converges to $\sqrt a$.

%
%
%
%
We extend the square root to its entire domain by repeatedly applying: 
\[
	\sqrt{x} = 2 * \sqrt{\frac{x}{4}}
\]

O'Connor's \Coq{} implementation~\cite{Oconnor:real} includes the much faster Newton iteration, whose complexity is logarithmic in the number of decimals. The function to iterate is:
\begin{lstlisting}
Definition f (r : Q) : Q := r / 2 + a / (2 * r).
\end{lstlisting}
Because of the absence of exact division on our approximate rationals we cannot implement this function directly. However, we can implement it on our real numbers. As the above definition does not use sharing, we have to define this function on the reals by first defining:
\begin{lstlisting}
Definition f (r : AQ) (ε : Qpos) : AQ := (r + approx_div (Qdlog2 ε) a r) ≪ (-1).
\end{lstlisting}
and then showing that it gives rise to a continuous function $f : \AQ \to \AR$ which we finally lift to a function $\bind\ f : \AR \to \AR$ on the reals. In this way we take care of sharing, division and intermediate use of the \lstinline|approx| function (see Section~\ref{section:reals}) all in one go. We hope the future correctness proof to be quite smooth, since we work with \emph{exact} real numbers. We have implemented this in \Haskell{} and it performs really well. 

\section{Benchmarks}\label{section:bench}
The first step in this research was to create a \Haskell{} prototype based on O'Connor's implementation of the real numbers in \Haskell~\cite{OConnor:mscs}. The second step was to implement and verify this prototype in \Coq. Our \Coq{} development contains verified versions of: the field operations, exponentiation by a natural, computation of power series, \exp, \atan, \sin, \cos, $\pi$ and the square root.

In this section we present some benchmarks, taken from the `Many Digits' friendly competition~\cite{niqui-wiedijk}, comparing the old and the new implementation, both in \Haskell{} and \Coq. All benchmarks have been carried out on an Intel Core Quad 2.4 GHz with 8GB of memory running \DebianFull{}. The sources of our developments can be found at~\url{https://github.com/c-corn/corn}.

\begin{table}[tp]
\centering
\begin{tabular}{|c|c|c|c|c|}
\hline
& Expression & \ \ Decimals \ \quad & \ \ Old \ \quad & \ \ New \ \quad  \\
\hline
P01 & $\sin\ (\sin\ (\sin\ 1))$ & 5.000 & 25s & 2.3s \\
P02 & $\sqrt{\pi}$ & 5.000 & 3.3s & 1.7s \\
P03 & $\sin\ \e$ & 5.000 & 13s & 1.2s \\
P04 & $\exp\ (\pi * \sqrt{163})$ & 5.000 & 22s & 2.0s \\
P05 & $\exp\ (\exp\ \e)$ & 5.000 & 43s & 2.6s \\
P06 & $\log\ (1 + \log\ (1 + \log\ (1 + \log\ (1 + \pi))))$ & 500 & 107s & 2.5s \\ 
P07 & $\exp\ 1000$ & 20.000 & 1.1s & 0.7s \\
P08 & $\cos\ (10^{50})$ & 20.000 & 6.7s & 1.4s \\
P09 & $\sin\ (3 * \frac{\log\ 640320}{\sqrt{163}})$ & 5.000 & 33s & 16s \\
P11 & $\tan\ \e + \atan\ \e + \tanh\ \e + \atanh\ \frac{1}{\e}$ & 500 & 41s & 3.2s \\
P12 & $\asin\ \frac{1}{\e} + \cosh\ \e + \asinh\ \e$ & 500 & 99s & 3.2s  \\
\hline
\end{tabular}
\caption{\Haskell{}, compiled with \texttt{ghc} version 6.12.1, using \texttt{-O2}. The column `old' refers to the \Haskell{} prototype of O'Connor, and the column `new' to our \Haskell{} prototype.}
\label{table:haskellO2}
\end{table}

\begin{table}[tp]
\centering
\begin{tabular}{|c|c||c|c|c||c|c|}
\hline
 & Expression & \ \ Decimals \ \quad & \ \ Old \ \quad & \ \ New \ \quad
 	& \ \ Decimals \ \quad & \ \ New \ \quad \\
\hline
P01 & $\sin\ (\sin\ (\sin\ 1))$ & 25 & 46s & 0.6s
	& 500 & 3.8s \\
P02 & $\sqrt\pi$ & 25 & 0.3s & 0.03s
	& 500 & 6.8s \\
P03 & $\sin\ \e$ & 25 & 36s & 0.1s
	& 500 & 1.9s \\
P04 & $\exp\ (\pi * \sqrt{163})$ & 10 & 214s & 0.1s
	& 500 &  3.7s \\
P05 & $\exp\ (\exp\ \e)$ & 10 & 36s & 0.2s
	& 500 &  3.2s \\
P07 & $\exp\ 1000$ & 10 & 2662s & 1.0s
	& 2.000 & 4.9s \\
P08 & $\cos\ (10^{50})$ & 25 & 11s & 0.3s
	& 2.000 & 12s \\
\hline
\end{tabular}
\caption{\Coq{} trunk, revision 14023. The column `old' refers to the \Coq{} implementation of O'Connor, and the column `new' to our \Coq{} implementation. Computations using a higher precision did not terminate within a reasonable time using O'Connor's implementation, so these are omitted.}
\label{table:coq}
\end{table}

Table~\ref{table:haskellO2} shows some benchmarks in \Haskell{} with compiler optimizations enabled (\texttt{-O2}) and Table~\ref{table:coq} compares our \Coq{} implementation with O'Connor's. More extensive benchmarking shows that our \Haskell{} implementation generally benefits from a 15 times speed up while the speed up in \Coq{} is generally more than a 100 times for small examples already. This difference between the comparison of the \Haskell{} and the \Coq{} implementation is explained by the fact that O'Connor's \Haskell{} implementation already uses rational numbers built from fast integers and incorporates various optimizations, while his \Coq{} implementation does not. The last column of Table~\ref{table:coq} indicates that our new implementation is able to compute an order of magnitude more decimals in the same amount of time.

We also compared the new reals built from \Coq{}'s fast rationals (Section~\ref{section:fast_rationals}) and our dyadic rationals (Section~\ref{section:dyadics}). For \exp, \sin{} and \cos{} we obtain quite similar results due to the our range reductions to reduce the length of the power series. In case of the square root, the dyadics rationals are much faster because wolfram iteration is designed for an efficient shift. It is interesting to notice that $\pi$ and \atan{} benefit the least from our improvements, as we are unaware of range reductions to reduce the length of the series.

We conclude this section with a comparison between the performance of Wolfram's algorithm in \Coq{} and \Haskell. The \Haskell{} prototype (without compiler optimizations) is quite fast, computing 10,000 iterations (giving 3,010 decimals) of $\sqrt2$ takes 0.2s. In \Coq{} it takes 7.4s using type classes and 7.2s without type classes. Here we exclude the time spend on type class resolution. Thus type classes cause only a 3\% performance penalty on computations, which is very acceptable for the modularity that they introduce.

Unfortunately, the \Coq{} implementation is slow compared to \Haskell. Laurent Théry suggested that this is due to the representation of the fast integers, which uses a tree with a fixed depth and when the size of the integer becomes too big uses a less optimal representation. Increasing the size of the tree representation and avoiding an inefficiency in the implementation of shifts reduces this time to 5.4s.

\section{Conclusions and Related work}
\label{section:conclusions}
We have greatly improved the performance of real number computation in \Coq{} using \Coq's new machine integers. We produced highly structured and abstract code using type classes with no apparent performance penalty. Moreover, \Coq's notation mechanism combined with unicode characters gives nicely readable statements and proofs. Type classes were a great help in our work. However, the current implementation of instance resolution is still experimental and at times too slow (at compile time).
 
Canonical structures provide an alternative, and partially complementary, implementation of type classes~\cite{adhoc}. By choice, canonical structures restrict to deterministic proof search, this makes them more efficient, but also somewhat more intricate to use. The use of canonical structures by the \Ssreflect{} team~\cite{packed} makes it plausible that with some effort we could have used canonical structures for our work instead. However, the \Ssreflect-library is currently not suited for setoids which are crucial to us. The integration of unification hints~\cite{Hints} into \Coq\ may allow a tighter integration of type classes and canonical structures.

We needed to adapt our correctness proofs to prevent the virtual machine from eagerly evaluating them. Lazy evaluation for \Prop\ would have allowed us to use the original proofs. Moreover, setoid rewriting over relations in \Type{} would have made our work much easier.

The experimental \lstinline|native_compute| by Boespflug, Dénès and Grégoire~\cite{native_compute} performs evaluation by compilation to native \OCaml{} code. This approach uses the \OCaml{} compiler available and is interesting for heavy compilation. Our first experiments indicate an additional speed up of 3 times compared to \lstinline|vm_compute|.

The \Flocq{} project~\cite{BolMel11} formalizes infinitary floating-points in \Coq. It provides a library of theorems on multi-radix multi-precision arithmetic and supports efficient numerical computations inside \Coq. However, the current library is still too limited for our purposes, but in the future it should be possible to show that they form an instance of our approximate rationals. This may allow us to gain some speed by taking advantage of fine grained algorithms instead of our more straightforward ones.

The encoding of real numbers as streams of `bits' is potentially interesting. However, currently there is a big difference in performance. The computation of 37 decimals of the square root of 1/2 by Newton iteration~\cite{JulienP09}, using the framework described in~\cite{bertot2007affine,julien2008certified}, took 12s. This should be compared with our use of the Wolfram iteration, which gives only linear convergence, but with which we nevertheless obtain 3,000 decimals in a similar time. On the other hand, the efficiency of $\pi$ in their framework is comparable with ours. Berger~\cite{berger2009coinductive}, too, uses co-induction for exact real computation.

The present work is part of a larger program to use constructive mathematics based on type theory as a programming language for exact analysis. This should culminate in a numerical ODE-solver. To do so we need to extend the current technology to functional analysis. For instance we will build a type class interface for metric spaces in order to treat various function spaces.

Cohen and Mahboubi~\cite{cohen:hal-00671809,COHEN:2012:INRIA-00593738:4} provide a formalization of quantifier elimination for the theory of decidable real closed fields, and an implementation of the algebraic real numbers. Quantifier elimination will automate many proofs in constructive analysis which only involve algebraic real numbers.
Conversely, our implementation could be used for efficiently evaluating a Cauchy representation of an algebraic real number.

\subsection*{Acknowledgements}
We thank Eelis van der Weegen for many discussions and Pierre Letouzey and Matthieu Sozeau for closing some of our bug reports. 
We are grateful to the anonymous referees who helped to improve the presentation of the paper.

\bibliographystyle{alphaabbr}
\bibliography{alg}
\end{document}